# Water in the Martian regolith from OMEGA/Mars Express.

Joachim Audouard, François Poulet, Mathieu Vincendon, Jean-Pierre Bibring, Brigitte Gondet and Yves Langevin
*Institut d'Astrophysique Spatiale (UPSUD/CNRS), Orsay, France.*

Ralph E. Milliken
*Dept. Geological Sciences, Brown University, Providence, RI, U.S.A.,*

Denis Jouglet
*CNES, Toulouse, France*

Corresponding author, email address and postal address:
Joachim Audouard
*joachim.audouard@ias.-psud.fr*
Institut d'Astrophysique Spatiale, Centre universitaire d'Orsay, Bâtiment 120–121, 91405 Orsay Cedex France

# ABSTRACT

Here we discuss one of the current reservoirs of water on Mars, the regolith and rocks exposed at the surface. This reservoir is characterized by the presence of $H_2O$- and OH-bearing phases that produce a broad absorption at a wavelength of ~3 µm in near-infrared (NIR) reflectance spectra. This absorption is present in every ice-free spectrum of the Martian surface obtained thus far by orbital NIR spectrometers. We present a quantitative analysis of the global distribution of the 3 µm absorption using the Observatoire pour la Minéralogie, l'Eau, les Glaces et l'Activité (OMEGA) imaging spectrometer that has been mapping the surface of Mars at kilometer scale for more than ten years. Based on laboratory reflectance spectra of a wide range of hydrous minerals and phases, we estimate a model-dependent water content of 4±1 wt. % in the equatorial and mid-latitudes. Surface hydration increases with latitude, with an asymmetry in water content between the northern and southern hemispheres. The surface hydration is compared to various parameters (albedo, dust, geological units, time, relative humidity, atmospheric water pressure, and in situ measurements performed by Phoenix and Curiosity) to constrain the nature of the reservoir. We conclude that the nature of the surface hydration of the Martian low latitudes is not adsorbed water but rather more tightly-bound water molecules and hydroxyl groups in the structure of the materials of the near-top surface. A frost-related process best explains the implementation of water into and onto the first microns of the high latitudes Martian regolith.

## 1. Introduction and background

Reflectance spectroscopy provides an accurate, reliable, non-destructive technique for identifying the presence of $H_2O$ or $OH^-$ in planetary surfaces and regolith to depth varying with the refractive indices (wavelength and composition) and with the regolith texture. In the near-infrared (NIR), the effective depth can vary to a few µm (for dust-like materials) to several cm (for clear $CO_2$ ice). In this paper, we refer to the effective surface sounded by NIR spectroscopy techniques as "optical surface" (typically the top ~100 µm of the regolith). On Mars, a strong absorption in the ~3 µm wavelength range was recognized by early telescopic and space observations and attributed to the presence of water-bearing materials (fundamental symmetric and asymmetric stretches) [*Houck et al.*, 1973; *Pimmentel et al.*, 1974; *Bibring et al.*, 1989; *Calvin*, 1997; *Yen et al.*, 1998; *Murchie et al.*, 2000]. An absorption near 3 µm was also reported in spectroscopic studies of the Moon [*Clark*, 2009; *Pieters et al.*, 2009; *Sunshine et al.*, 2009; *Izawa et al.*, 2014] and of many asteroids [*Rivkin et al.*, 2013; *Beck et al.*, 2014], as well as at Vesta's surface [*De Sanctis et al.*, 2012]. This absorption is sometime referred to as "water-induced", 3 µm "water", or "hydration" absorption. Though $H_2O$ molecules are often a dominant contributor to this broad absorption, $H_2O$-free materials can also exhibit absorptions near 3 µm due to the presence of OH- [e.g. *Milliken and Mustard*, 2005]. More recently, orbital measurements by imaging spectrometers in this spectral range have allowed refinement of the 3 µm signature at local, regional, and global scales [*Jouglet et al.*, 2007; *Milliken et al.*, 2007]. These previous analyses were performed from the Observatoire pour la Minéralogie, l'Eau, les Glaces et l'Activité (OMEGA) data set taken during one Martian year, and they revealed general trends in the distribution Martian surface hydration as well as possible spatial and temporal variations in water content possibly linked to surface composition (albedo) and season (solar longitude, $L_s$). In parallel to these observations, our knowledge of the physical properties of the water molecules in a hydrated phase has been improved by theoretical [*Möhlmann*, 2008] and experimental studies [e.g. *Anderson et al.*, 1967; *Fanale and Cannon*, 1974; *Zent and Quinn*, 1997; *Jänchen et al.*, 2006, 2008; *Pommerol et al.*, 2009; *Beck et al.*, 2010], whereas modeling [e.g. *Jakosky*, 1983b; *Zent et al.*, 1993] and numerical global simulations [*Houben et al.*, 1997; *Böttger et al.*, 2005] have proposed that the surface hydration may play an important role on the Martian water cycle, especially in relation to the observed diurnal variation of atmospheric water vapour [*Farmer et al.*, 1977; *Titov*, 2002; *Smith*, 2002; *Melchiorri et al.*, 2009; *Maltiagliati et al.*, 2011].

Fundamental and overtone vibrational absorptions of $H_2O$ and $OH^-$ observed in reflectance spectra in the 1.3-5.0 µm wavelength region can be used to detect the surface hydration and to estimate the absolute or relative reservoir in particulate materials. Absorptions in the 3 µm wavelength region may be caused by several forms of $OH^-$ and/or $H_2O$, including hydrous minerals, water ice, or surface-adsorbed species (see a review by Dyar et al. [2010] and references therein). Different classes of $OH^-$ and $H_2O$ have been discriminated based on the mobility of the molecules: $OH^-$ and $H_2O$ can be "structural", i.e. components of a mineral ($OH^-$ is in the fundamental formula of many mineral species and $H_2O$ can be solvated in the structure of hydrous minerals such as in smectite clays), "bound" i.e. chemisorbed (covalently bound to metal ions and defects in minerals), or, at lower energies, "adsorbed" i.e. physisorbed (bound only by Van der Walls electrostatic forces thus very labile) where it can form a liquid-like film by hydrogen bounding. Overall, the strength of the mineral-water bound depends on the ion availability, 3-D porous structure and on the metal type of the mineral. Liquid $H_2O$ presents a broad absorption between 2.8 and 3.7 µm with a maximum near ~2.95 µm. Crystalline $H_2O$ ice absorption is narrower, with a maximum of at ~3.1 µm [Pommerol et al., 2009], whereas hydroxylation (i.e. covalently bound $OH^-$) of anhydrous material has been reported as responsible for an absorption near ~2.8 µm in lunar surface spectra, where soil interaction with solar wind is strong [*Managadze et al.*, 2011; *Ichimura et al.*, 2012]. Such factors are much weaker on Mars, and it has been suggested that minerals at the surface transiently adsorb atmospheric water vapor and thus the regolith could be a potential source/sink of $H_2O$, in line with observed diurnal variations of atmospheric water vapour [*Jakosky*, 1983a, 1985; *Zent et al.*, 1986; *Möhlmann*, 2008; *Pommerol et al.*, 2009; *Beck et al.*, 2010; *Temppari et al.*, 2010; *Maltiagliati et al.*, 2013]. The overlap of the $OH^-$ and $H_2O$ absorptions in the 3 µm region makes it difficult to distinguish between these two contributors but the absorption peak seen in ice-free Martian spectra in this spectral region is between 2.9 and 3.0 µm [Jouglet et al., 2007; Milliken et al., 2007], indicating that the dominant contributor is likely $H_2O$ without excluding potential $OH^-$ presence as well. In previous analysis of OMEGA data, *Jouglet et al.* [2007] and *Milliken et al.* [2007] both reported a seasonal variation of the surface hydration on Mars at high latitudes, suggesting that at least a portion of the observed surface hydration is mobile.

However, several properties of a given material can affect the shape and strength of the 3 µm absorption (e.g., albedo, particle size, and composition/crystal structure). Initial work by *Yen et al.* [1998] and *Cooper and Mustard* [1999] showed that these effects were important to consider for quantitative analyses of spectra. The understanding of these effects were further analysed and partly quantified by various studies [e.g., *Milliken and Mustard*, 2005, 2007a, 2007b; *Pommerol and Schmitt*, 2008a], yielding relationships between absorption strength and particle size, albedo, and composition. Observation conditions may also play a role when studying these spectral features with remote sensing techniques. In this respect, the potential observational biases induced by surface photometric effects have been documented by *Pommerol and Schmitt* [2008b]. In addition, remote observations of Martian surface materials are performed through an atmosphere dense enough to modify apparent reflectance spectra, and the precise impact of the atmospheric volatiles remains poorly characterized.

In situ experiments performed by Phoenix and Curiosity allowed a quantitative measure (in temperature) of the energy of the $OH^-$ and $H_2O$ bounding to the material and thus place constraints on the hydration nature of the Martian surface. Recent SAM- (Sample Analysis at Mars) and CheMin- (Chemistry and Mineralogy) based results from measurements of Martian regolith at Gale Crater (near the equator) have shown that most of the hydration observed in the heating experiments is likely not adsorbed [*Leshin et al.*, 2013], but rather water that is part of an amorphous phase and that is held at higher energies [*Bish et al.*, 2013]. The ChemCam (Chemistry & Camera) instrument measured no variation of the H LIBS signal with daytime and depth [*Meslin et al.*, 2013], indicating that H was strongly bound to their hosts and present in similar abundance in the bulk and at the top surface of the regolith. *Meslin et al.* [2013] then proposed that the variations of hydration seen from orbit could reflect the abundance of the amorphous component at the surface. Water ice was observed a few cm below the surface at the Phoenix landing site [*Mellon et al.*, 2011], and *Arvidson et al.* [2009] proposed that the mechanical cohesiveness of Phoenix regolith was controlled by the quantity of adsorbed water and water ice within the regolith above the ice table. The TECP dielectrical experiment onboard the Phoenix lander measured an important signal interpreted as high quantity of adsorbed water within the regolith [*Stillman and Grimm*, 2011] and a mechanism involving a thin film of liquid-like water (in the sense of *Möhlmann* [2008]) was proposed by *Cull et al.* [2010] to explain the vertical distribution of perchlorates in the vicinity of Phoenix.

The present work aims at re-evaluating the hydration state of the Martian surface using a more complete OMEGA dataset to quantify the water content of the optical surface of Mars. The goal is to place better constraints on the nature of $OH^-$ and $H_2O$ in the Martian regolith and to evaluate its possible role in the modern hydrological cycle. After a presentation of the dataset and hydration monitoring method in Section 2, a global map is presented in Section 3 where spatial and seasonal variations are examined. Finally, we discuss in Section 4 the relationship of the $H_2O$ and OH bearing phases with various parameters including other water reservoirs to constrain the nature of the observed surface hydration.

## 2. Dataset and method

The hydration of the Martian surface is inferred from spectroscopic data obtained by OMEGA onboard Mars Express (MEX). This instrument records measurements of the radiance coming from the martian surface in the spectral range 0.36 – 5.1 µm divided into 352 spectral elements ("spectels") over three detectors: the visible (VIS) channel that is made of 96 spectels ranging from 0.36 to 1.07 µm; the C- and the L-channels are each composed of 128 spectels, ranging respectively from 0.93 to 2.7 µm and from 2.53 to 5.1 µm [*Bibring et al.*, 2004]. The hydration analysis based on the 3 µm signature requires careful data selection criteria and processing methods that are described in detail in this section.

## 2.1 Data

This study of the 3 µm hydration feature requires use of both the C and L channels, leading us to select OMEGA datacubes when both detectors are turned on. The L detector must be cooled at 78 K and data acquired with higher detector temperature are excluded from this study. We restrict our analysis to data with solar incidence angle (*i*) less than 85° and emergence angle (*e*) lower than 20°. Data taken during the 2007 global dust storm are also excluded. As extensively described in *Jouglet et al.* [2007; 2009], the L channel is subject to variations of its radiometric performance. It is established that the radiance measured by the L channel depends on the calibration lamp level, which is measured at the beginning of every orbit. An empirical correction for the non-nominal orbits was implemented by *Jouglet et al.* [2009]. *Audouard et al.* [2014] addressed the reliability of this correction and concluded that the empirical correction enabled the use of non-nominal orbits for scientific studies. In total, 6132 out of 9643 OMEGA data cubes (up to orbit #8485, after which the SWIR-C detector was shut down) fulfill these conditions and are processed for the present study. This data set corresponds to approximately 350 million spectra, spanning four full Martian years. Additional selections based on atmospheric conditions and on data signal to noise ratio (SNR) are discussed in Sections 2.3 and 2.4.

## 2.2 Spectroscopic monitoring of hydration

OMEGA spectra are corrected for atmospheric attenuation through division by a power-scaled transmission spectrum of the Martian atmosphere, similar to the method presented in *Langevin et al.* [2007]. Contributions from thermal emission are non-negligible at wavelengths >3 µm during the daytime, and this effect is removed using the temperature retrieval method described in *Audouard et al.* [2014]. The radiance spectra from 1 to 4 µm are then converted to reflectance factor and then to single scattering albedo ($\omega$) using *Hapke* [1993] theory, as described in *Milliken and Mustard* [2005]. Next steps consist of dividing every spectrum by a linear continuum derived from the single scattering albedo values at 2.35 and 3.7 µm, and computing the effective single-particle absorption-thickness (ESPAT parameter) at 2.9 µm as described in *Milliken and Mustard* [2007a] i.e. assuming isotropic phase function ($p(g) = 1$) and no opposition effect ($B(g) = 0$). These assumptions are likely false for the Martian surface at least at the macroscopic scale [e.g. *Vincendon, 2013*], and might therefore introduce a photometric bias in our results as shown by laboratory measurements [*Pommerol and Schmitt, 2008a*]. However, given the incidence and emergence angles of the selected dataset, the photometric bias is expected to remain small [*A. Pommerol*, personal communication]. We have checked for a dependency of the ESPAT parameter at 2.9 µm on the phase angle but found no significative trend, indicating that this effect is small for the selected dataset. This effect is therefore not discussed further in this article. The ESPAT parameter at 2.9 µm has been related to the water content of various materials thanks to controlled sorption-desorption experiments of relevant hydrous minerals and phases under controlled laboratory environments [*Milliken and Mustard*, 2005, 2007a, 2007b; *Pommerol and Schmitt*, 2008a, 2008b]. Specifically, the water weight percent (water wt. % hereafter) of many Martian regolith analogues can be approximated by a linear function: *water wt. % = A \* ESPAT(2.9 µm)*, where the value of *A* may vary depending on the material's composition and grain size, but with the advantage of being independent of the albedo [*Milliken*, 2006; *Milliken et al.*, 2007a; *Pommerol and Schmitt*, 2008b]. A typical linear coefficient $A = 4.17$ is derived by *Milliken et al.* [2007a, Figure 17] from a fit to sieved (< 45 µm) palagonite and montmorillonite data with various admixtures of darkening agents, with a residual 1σ standard deviation of ~1.15 water wt. %. Two different linear fits to the experimental data of *Milliken et al.* [2007a, Figure 17] give values of A = 4.00 and A = 4.29, when considering only data with respectively "large" or "small" sizes of darkening agent grain (but with <45 µm sized brighter hydrated materials). In the following analysis of the Martian surface water content we thus take A = 4.17 ± 0.15 to represent intermediate particle sizes for dark components mixed with brighter, <45 µm sized hydrated phases.

Although fine grains are expected to dominate the spectral signature of a mixture of various grain sized materials [*Milliken et al.*, 2007], the Martian surface exhibits a large particle size range from micrometer-sized particles ('dust') to coarse grain sizes ( >1mm) [*Palluconi and Kieffer*, 1981; *Mellon et al.*, 2000]. When considered alone (without any mixing with smaller grains), large-size particles exhibit much lower values of *A* than fine-grained samples, possibly as low as 0.6 for 250 µm sized materials [*Milliken and Mustard*, 2007b]. The major assumption of the ESPAT technique as applied to the Martian regolith in this study is thus the reliance on a single grain size that is meant to represent the mean size of the *spectrally* dominant components of the optical surface. The potential impact of this dependence as well a possible albedo-dependent bias will be discussed in section 3.1.2.

However, because we have restricted our analysis to data recorded with low (< 20 °) emergence angles, most of the OMEGA acquisitions with high phase angles are de facto excluded from the analysed data set.. We checked a possible dependence but no significant trend has been detected and it is not discussed further here.

## 2.3. Impact of surface and atmospheric volatiles

Water ice (as clouds in the atmosphere or as surface frost) significantly affects the radiance observed from orbit in the 3 µm spectral region [e.g. *Vincendon et al.*, 2011; *Madeleine et al.*, 2012]. As we aim to quantify the hydration of ice-free surfaces, the 1.5 µm band depth described in *Langevin et al.* [2007] is used to detect icy frost presence at the surface and thick clouds in the atmosphere. Every OMEGA pixel with a band depth > 1.5% is thus excluded from the analysis. However, not all water ice clouds will be excluded with this threshold because the absorption at 1.5 µm decreases for thin clouds or clouds with very small particles. Water ice clouds can also be identified using another ratio criterion (reflectance at 3.4 µm divided by reflectance at 3.52 µm), first implemented by *Langevin et al.*, [2007]. This water ice cloud index is calculated for the entire selected dataset. Its values range from 0.5 to 1.0 (Figure 1). The smallest values unambiguously indicate the presence of icy clouds. The latitudinal and seasonal distribution of water ice clouds are easily detected by this parameter (Figure 1a), and they match well with those derived from the TES/MGS instrument [*Smith*, 2004]. Figure 1b presents the distribution of derived water wt. % values as a function of this ice cloud index, and it is clear that increased presence of water ice clouds yields higher 'surface' hydration estimates. *Madeleine et al.* [2012] proposed a threshold of 0.8 for this index as a proxy for water ice cloud detection. We adopt this value for our study and exclude any OMEGA pixel with a value < 0.8 (additionally to the water ice 1.5µm filtering). It should be kept in mind that even with this threshold a contribution from thin water ice clouds or small-grained water ice haze may still be present in some observations.

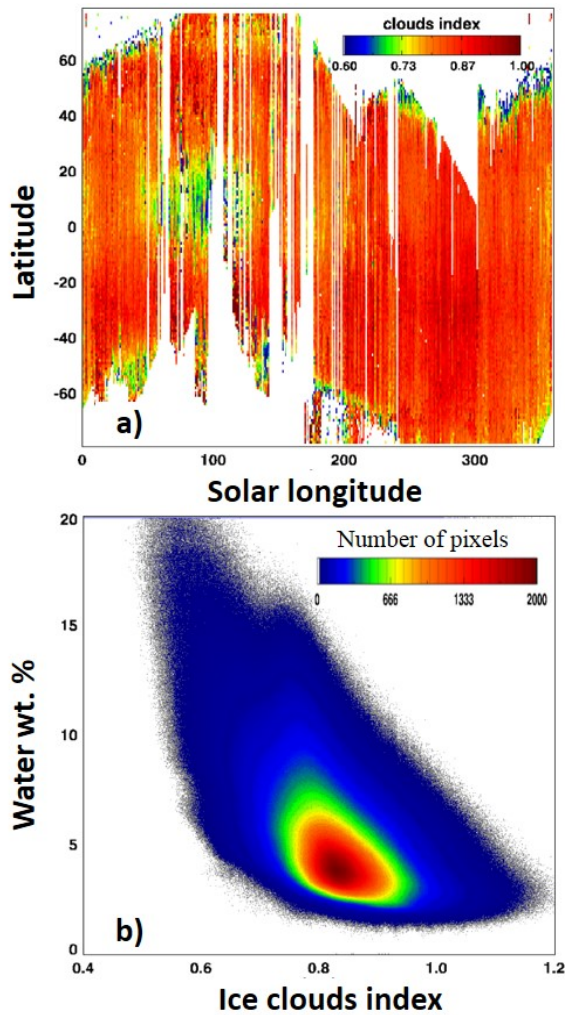

*Figure 1. a) Map of the water ice clouds index for the selected dataset as a function of solar longitude (Ls) and latitude. Lower values indicate higher water ice clouds opacity. The aphelion cloud belt and the high latitude water ice clouds are clearly visible on this map in blue, green and yellow. b) 2-D density histogram of computed water wt. % as a function of the water ice clouds index. Spectra showing a water ice signature at 1.5 µm are not present in these diagrams. See text for discussion.*

Martian airborne dust also disturbs remote observations of the surface in the 3 µm spectral region [*Vincendon et al.*, 2009]. In that range, Martian aerosols scatter light and also exhibit a 3 µm absorption feature similar to other martian material [*Määttänen et al.*, 2009], which modifies the apparent surface properties. The precise impact of airborne dust at 3 µm is difficult to predict because of variations in the amount and size of aerosols with season and location (latitude, longitude). For every OMEGA pixel we evaluate the effective dust opacity, $\tau_{eff}$, as $\tau_{eff} = \tau_{MER} / \cos i$, where $i$ is the solar incidence angle and $\tau_{MER}$ is the dust opacity measured by the Mars Exploration Rovers (MERs) as a function of solar longitude [*Lemmon et al.*, 2014]. $\tau_{MER}$ is scaled to the local pressure using the pressure predictor of *Forget et al.* [2007], as detailed in *Audouard et al.* [2014]. At latitudes lower than ~60°, retrievals of dust opacity with OMEGA are in fair agreement with $\tau_{MER}$ (within ±20%, *Vincendon et al.*, [2009]). We thus consider $\tau_{MER}$ as a reliable proxy for the atmospheric dust load at a given Ls and we use $\tau_{eff}$ to estimate the amount of aerosols contributing to each OMEGA observation.

In order to evaluate the impact of dust opacity on the water wt. % retrieval using the ESPAT parameterization, we perform a water content mapping of the Mawrth Vallis region from observations taken at different dust opacities (Figure 2). Mawrth Vallis presents the largest abundance of phyllosilicates on Mars at the kilometre scale [*Poulet et al.*, 2005, 2008] and previous investigations of the 3 µm absorption have already revealed a 3 µm absorption over the phyllosilicate-bearing terrains that is stronger than adjacent terrains [*Jouglet et al.*, 2007; *Milliken et al.*, 2007]. As illustrated in Figure 2, the overall hydration decreases with increasing $\tau_{eff}$. For instance, the maximum of water wt. % found on the phyllosilicate-bearing terrains decreases from 9% during clear atmosphere (Figure 2a) to 3.0% during aerosol-loaded atmosphere (Figure 2e). Aerosols blur and can obscure the apparent 3 µm absorption of the Martian surface. Unless a comprehensive aerosol model (real aerosol sizes, real $\tau_{eff}$, and precise effect of the aerosols at 3 µm) is taken into account to remove their contribution, the hydration of the surface is always biased from orbital measurements under dusty atmospheric conditions, though observations acquired under the clearest atmospheric conditions will provide the most reliable estimate of true surface hydration. A specific data filter has to be implemented to exclude observations with significant atmospheric aerosol loads. In this respect, several threshold value on $\tau_{eff}$, between 2 and 3, were tested and it appears that a threshold of 2.5 is the best trade-off between removal of most of the dust-contaminated observations and keeping valuable data and spatial coverage.

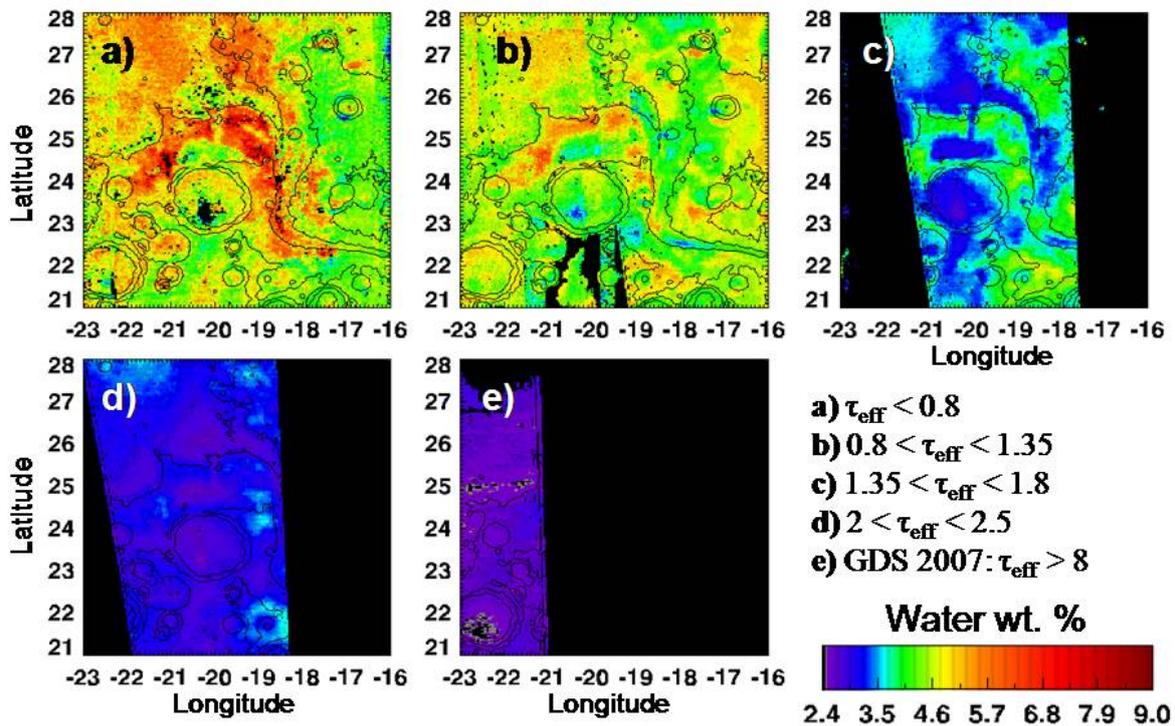

*Figure 2. 32 pixel per degree (~2\*2 km / pixel) maps of water content over the Mawrth Vallis region, where large phyllosilicate-bearing terrains are present. Five maps are built using different $\tau_{eff}$ opacity ranges. All data were previously filtered using a water ice clouds index greater than 0.8. The airborne dust blurs and lowers the apparent water content of the surface whose mineralogical contrasts are barely visible at $\tau_{eff} = 2$.*

Finally, water vapour in the atmosphere could potentially have an impact on the water-related 3 µm absorption, as water vapor lines, although sharp and narrow, are superposed on the larger 3 µm band. We have estimated the water vapour concentration in every OMEGA pixel using its absorption at 2.65 µm [*Melchiorri et al.*, 2007] but found no correlation with our water wt. % values.

## 2.4. Impact of data SNR

As described above, the ESPAT parameter is computed from the single-scattering albedo, ω, at 2.9 µm divided by the continuum between 2.35 and 3.7 µm. The ω calculation will be affected by the OMEGA instrumental noise, dominated by the read noise equal to 1.85 Digital Units for every wavelength [*Bonello et al.*, 2005]. The error on the water content resulting from the instrumental noise is then defined as the variation of water content when the values of the continuum and the ω at 2.9 µm are shifted in opposite directions due to noise. This evaluation is performed for one typical bright region (Tharsis) and one typical dark region (Syrtis Major). For these two terrains, the maximum error induced by data SNR onto retrieved water wt. % values is shown in Figure 3 as a function of SNR. The error decreases exponentially with SNR as $\Delta \text{water wt.\%} = \frac{\text{water wt.\%}}{100} * e^{5.33 \pm 0.16 - 0.97 \pm 0.01 * \ln(\text{SNR})}$ for both locations. We restrict our study to data that lead to a maximal error of 0.5%, which corresponds to SNR greater than 40. Figure 3b is a cumulative frequency histogram of the data set as a function of SNR. Of the selected ~350 millions of spectra, 76% (i.e. ~270 millions of spectra) have a SNR greater or equal to 40, thus with water wt. % relative uncertainty lower than 0.5 wt. %.

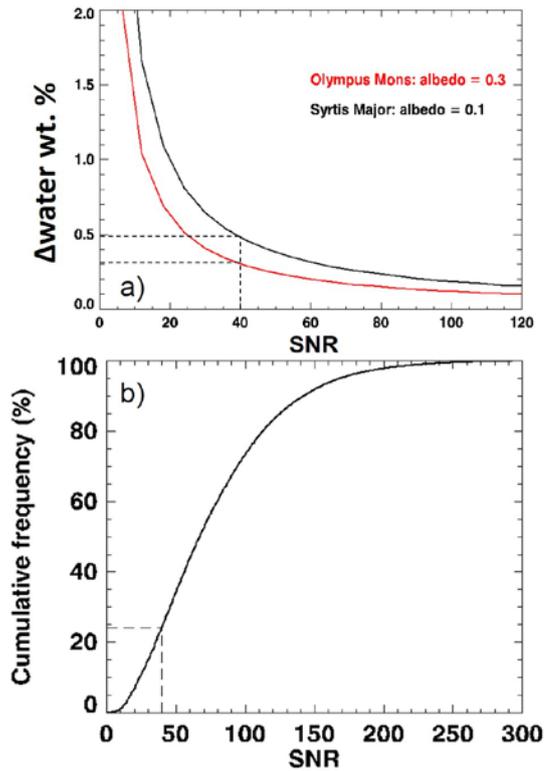

*Figure 3. a) Error on water wt. % values induced by OMEGA SNR in the 3 μm spectral region; the estimate of the error is shown for bright and low albedo regions. b) Distribution of SNR in the 3 μm spectral region for the selected dataset. The dashed line indicates the SNR threshold retained for the filtering process, yielding to an error on water wt. % of ~0.5 wt. % for dark terrains and ~0.3 wt. % for bright terrains.*

## 2.5. Uncertainties

Before applying the ESPAT technique to the selected OMEGA data, we summarize the uncertainties on the inferred water content. These uncertainties have three sources: uncertainties on the chosen ESPAT-H$_2$O relationship, the OMEGA data, and the observational conditions. As mentioned previously, we use the relationship *water wt. % = A \* ESPAT(2.9 μm)* with $A = 4.17 \pm 0.15$. The 0.15 % of ESPAT uncertainty is within the residual 1σ standard deviation of ~1.15 water wt. % resulting from compositional and albedo-dependent uncertainty [*Milliken* *and Mustard*, 2007b], which can be directly propagated to the absolute water wt. % value. An additional source of uncertainty due to the methodology comes from the assumed grain size distribution of particle size (spectrally dominant components are < 45 µm in diameter). Values as low as A = 0.6 were derived by *Milliken and Mustard* [2007b] for large grain sizes (> 250 µm), potentially leading to very large errors (up to a factor 7) on the derived water wt. %. Although the assumed grain size estimate may be wrong on a local scale, the error is considered to be systematic for this study because most kilometre-scale pixels will have significant components of regolith when evaluating trends at the global scale (that is, at a global scale the surface is dominated by regolith materials with components <45 µm in diameter) [*Milliken et al.*, 2007b]. It is however important to keep in mind that this grain size effect could imply a systematic overestimation of the water content presented here if the spectrally dominant components have an average grain size >45 µm.

The major relative uncertainty on the water wt. % values comes from increased dust opacity that can act to reduce the strength of the 3 um band and thus the derived water content. As mentioned in section 2.3, a filtering processing allows rejecting the observations that are strongly blurred by the dust opacity, and thus limiting the local and time variations due to variable dust opacities. However, there is still a likely systematic underestimation of water content as derived from orbital measurements (see Figure 2). Finally, there is a specific selection of the data that limits the relative error on the water content due to OMEGA SNR to 0.5%.

In summary, when observing a given location on Mars, systematic errors in the ESPAT to water wt. % relationship arise from compositional (1σ = 1.15 wt. %) and grain size (possibly up to a factor 7) uncertainties. Relative errors on water wt. % are caused by data SNR (< 0.5 wt. %) and atmospheric volatiles content during the observation. Some of these factors can produce large errors, so that any comparison with in situ measurements should be used with caution. However, most of these errors are systematic, allowing us to confidently examine global trends using a unique ESPAT-water wt. % linear relationship A global scale comparison between the water content and thermal inertia will be performed (Section 3.1.2) to check any size dependence, and the impact of aerosols will be evaluated using the $\tau_{eff}$ proxy in section 3.

# 3. Results

In this section, we first present a global map of water wt. % using the most complete data set to date and discuss the correlations with other global surface properties. Evolution of water wt. % values with time (season and day) are then presented and we finally compare our retrievals with several *in situ* measurements of water content.

## 3.1. Spatial variations of the water content

### 3.1.1 Global distribution

Figure 4 is a 32 pixel per degree global map of mean water wt. % derived from the selected OMEGA dataset and using the relationship *water wt. % = 4.17 ESPAT$_{2.9\mu m}$*. Pixels with ice clouds index lower than 0.8, water ice absorption at 1.5 µm greater than 1.5% and effective dust opacity greater than 2.5 were excluded from the dataset used to build this map. The overall trend in the spatial distribution of surface hydration is consistent with previous mapping [*Milliken et al.*, 2007], though our results provide significantly improved coverage. Most of the discrepancies observed between these results and previous results are attributable to the filtering processes applied in this work that exclude data containing thin water ice clouds, which was not the case in *Milliken et al.* [2007]. Some water ice clouds may be still present on Figure 4 (near 70°N of latitude), but they are barely visible at a global scale. The cloud index threshold that we used could be more restrictive but it would be at the cost of decreased spatial coverage.

It is clear from Figure 4 that the water wt. % is correlated primarily with latitude. Little variation in surface hydration is seen at latitudes lower than 30°, but water content increases dramatically at latitudes poleward of ±30°. This latitudinal variation, especially for the northern hemisphere, is unambiguous and well beyond the uncertainties described above, especially given that calculated water content at latitudes of ~70°N is twice that observed at equatorial latitudes. The latitudinal increase in the north begins to become strong at ~40°N. In the southern hemisphere, the increase is about half that of the Northern hemisphere but is still apparent for any season beginning at ~30°S (see next Section).

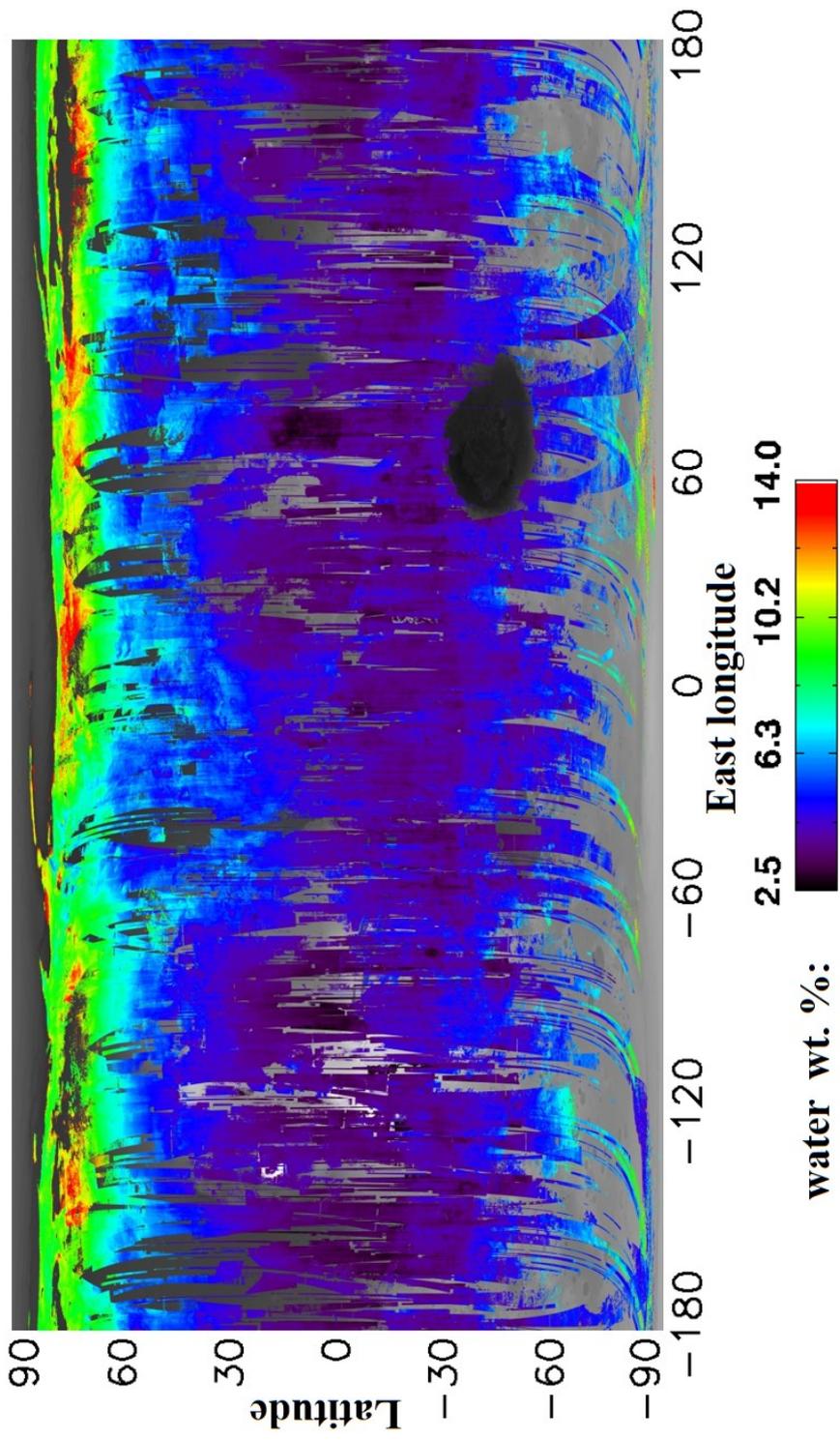

*Figure 4. Global map of mean water wt. % at a resolution of 32 pixel per degree (i.e. ~1.8 × 1.8 km at the equator) in cylindrical projection. Data is from all season excluding pixels contaminated by water ice at the surface, water ice clouds and tau > 2.5. The background image is MOLA and ~68.5% of the bins are filled.*

### 3.1.2 Variations with surface properties

#### 3.1.2.1 Thermal inertia

The global distribution of water wt. % exhibits no correlation with published thermal inertia of the Martian surface [*Putzig and Mellon*, 2007; *Audouard et al.*, 2014]. Low thermal inertia (e.g. Tharsis) and high thermal inertia terrains (e.g. Syris Major) have comparable water wt. % despite their large difference in bulk grain size inferred from thermal inertia. Without excluding potential systematic grain size bias, the <45 µm fraction of the surface material could dominate the spectroscopic features in the 3 µm region, as proposed by *Milliken et al.* [2007].

#### 3.1.2.2 Albedo

The variation of water content at latitudes lower than 30° as a function of albedo is shown in Figure 5a. After the filtering process, 82 % of the equatorial to mid-latitudes compose this 2D histogram; their water wt. % values remain constant with latitude with a mean value of ~4.0 water wt. % and a 3σ of ~1.7 water wt. %. We remark that the bright and dark terrains exhibit lower average values than terrains with intermediate albedo. This difference could indicate variations in surface composition and/or grain size. However, the discrepancy is well within the compositional uncertainties on the water wt. % values (1σ = 1.15 wt. %, see section 2.5), indicating that our water wt. % retrieval method is effectively not biased by the albedo of the surface, similar to results of *Milliken et al.* [2007].

At high latitudes, the longitudinal variations of water wt. % are positively correlated with the abundance of dust at the surface. Dust abundance is evaluated with OMEGA using the absorption feature of nanophase ferric oxide centered at 0.86 µm [*Ody et al.*, 2012]. Figure 5b shows this correlation for the northern high latitudes: dusty surfaces have a significantly stronger hydration in comparison to less dusty surfaces. Although the spatial coverage over the southern high latitudes is more limited because of the higher airborne dust opacity during southern summer, a similar but slightly weaker (by several wt. %) positive trend is observed in the southern hemisphere.

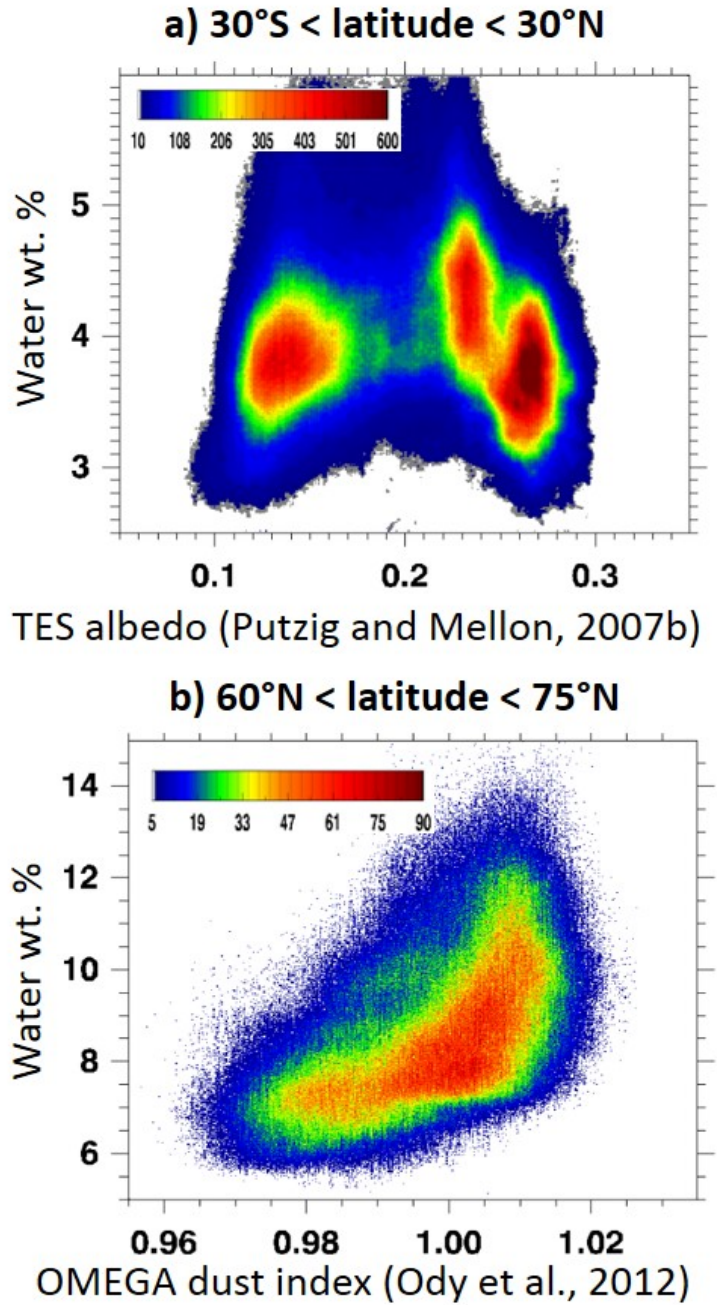

*Figure 5. 2D density histogram where colours indicate the number of pixel falling into every bin : a) Values of water wt. % for latitudes between 30°S and 30°N as a function of albedo derived from TES [Putzig and Mellon, 2007]; b) values of water wt. % for latitudes between 60°N and 75°N as a function of dust index derived from OMEGA [Ody et al., 2012]. Note the different y-axis in the two diagrams.*

3.1.2.3 Composition

Some large local variations (up to 5 water wt. %) with compositional properties of the surface exist as reported for the phyllosilicate- and sulphate-bearing deposits in Mawrth Vallis and Terra Meridiani [*Jouglet et al.,* 2007]. *Jensen and Glotch* [2011] noticed that the chloride-bearing deposits exhibit a smaller 3 µm signature with CRISM, possibly indicative of a smaller water content. Figure 6, which shows a chloride-bearing terrain, is consistent with such a trend, for which a decrease of ~1 wt. % $H_2O$ is observed in the chloride deposit in comparison to surrounding terrains. Although this difference is within the uncertainties, the excellent spatial correlation between the water distribution and the outlines of chloride deposit is in favour of composition-related variation.

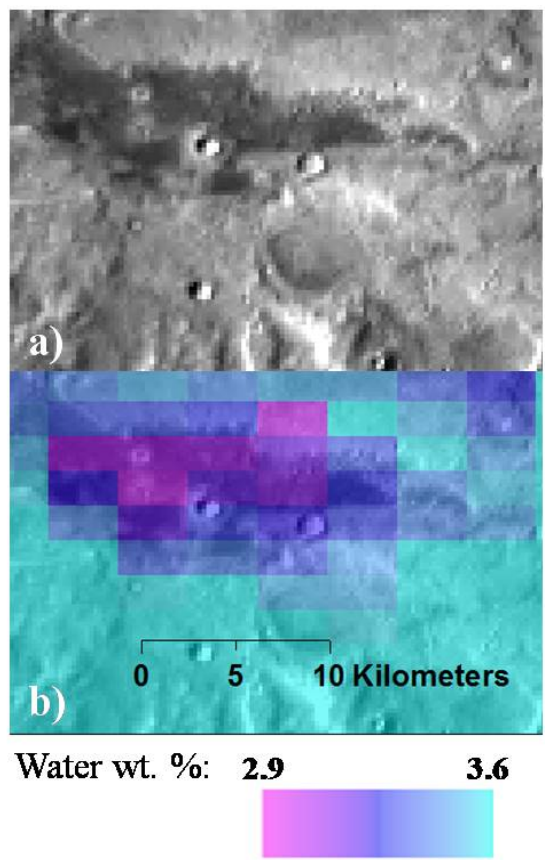

*Figure 6. a) THEMIS infrared daytime mosaic centered at [-139°E, -39°N].* Oosterlo et al. *[2010] proposed that the dark feature is chlorides-bearing material. b) map of mean water*

*wt. %. As expected and already qualitatively observed by CRISM [Jensen and Glotch, 2011], chloride-bearing terrains are dessicated relatively to surrounding terrains.*

The spatial distribution of surface hydration around the North Pole exhibits several variations. Figure 7 presents three polar maps: Figure 7a is the bolometric albedo derived from the TES instrument [*Putzig and Mellon*, 2007], Figure 7b is the mean water wt. % derived from OMEGA (same as Figure 4) and Figure 7c is the water-equivalent hydrogen derived from GRS [*Maurice et al., 2011*]. At constant latitude, longitudinal variations of OMEGA water wt. % (Figure 7b) are clearly observed. A lower surface hydration near 75°N, 50°W can be identified. This area corresponds to terrains with low water equivalent hydrogen as measured by GRS (Figure 7c). In this case, this suggests a possible local relationship of unknown origin between lack of buried ice and the surface water (discussed in section 4.2).

The circum-polar low albedo dune field (located between ~75°N and ~82°N) has a lower hydration (~9 wt. %) than other terrains (>11%) at similar latitude. This could be the result of a specific variation in material grain size or composition. Variable abundances of calcium sulfates and other hydrated minerals have been identified as part of the dune material [*Langevin et al.*, 2005], but the observed water wt. % distribution does not match the known spatial distribution of areas with large sulphate abundance (between 135° and 180°E). These dunes are composed of mobile material [*Massé et al.*, 2012] and have been observed to migrate up to several meters per year [*Hansen et al.*, 2011]. The different apparent hydration level of the dune field could thus reflect a grain size distribution different from the surrounding dust-bearing terrains. In addition, the apparent hydration variations within the circum-polar dune field could be an effect of sand sorting. Dunes composed of larger grains are likely to be less mobile and would be expected to have a lower value of the *A* coefficient. GRS does not observe any variation of subsurface hydrogen in the circum-polar dune field nor between the dune fields, possibly due to the large (~600 km) footprint of those data.

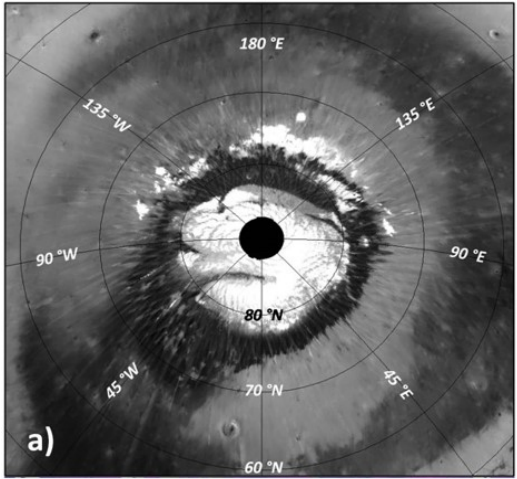

0.06 **Albedo** 0.32

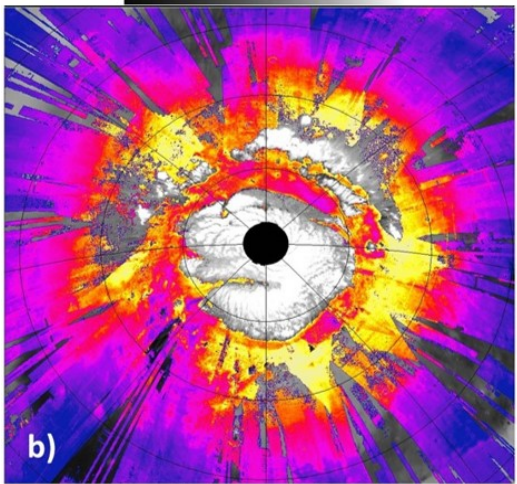

Water wt. %
5  7  9  11  13

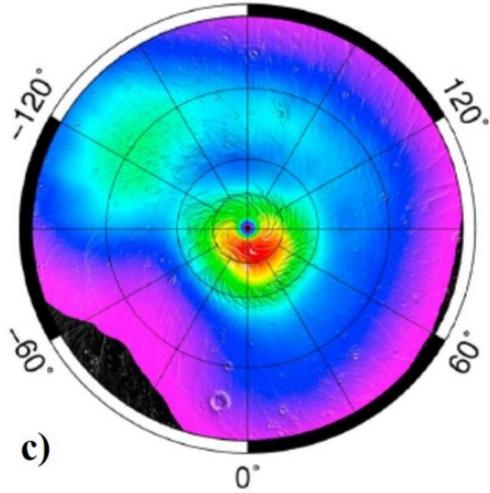

**GRS frost-free WEH in % (modified from Maurice et al., 2011)**

10  23  36  49  62  75

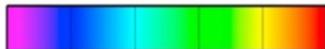

*Figure 7. Lambertian cyclindrical projections of the north pole region of: a) bolometric albedo measured by TES [Putzig and Mellon, 2007] ; b) OMEGA-derived water wt. % and c) GRS-derived WEH map polewards of 60°N latitude [Maurice et al., 2011]. The same minimum of hydrogen-bearing phase is seen near [70°N, 45°W] by both GRS and OMEGA.*

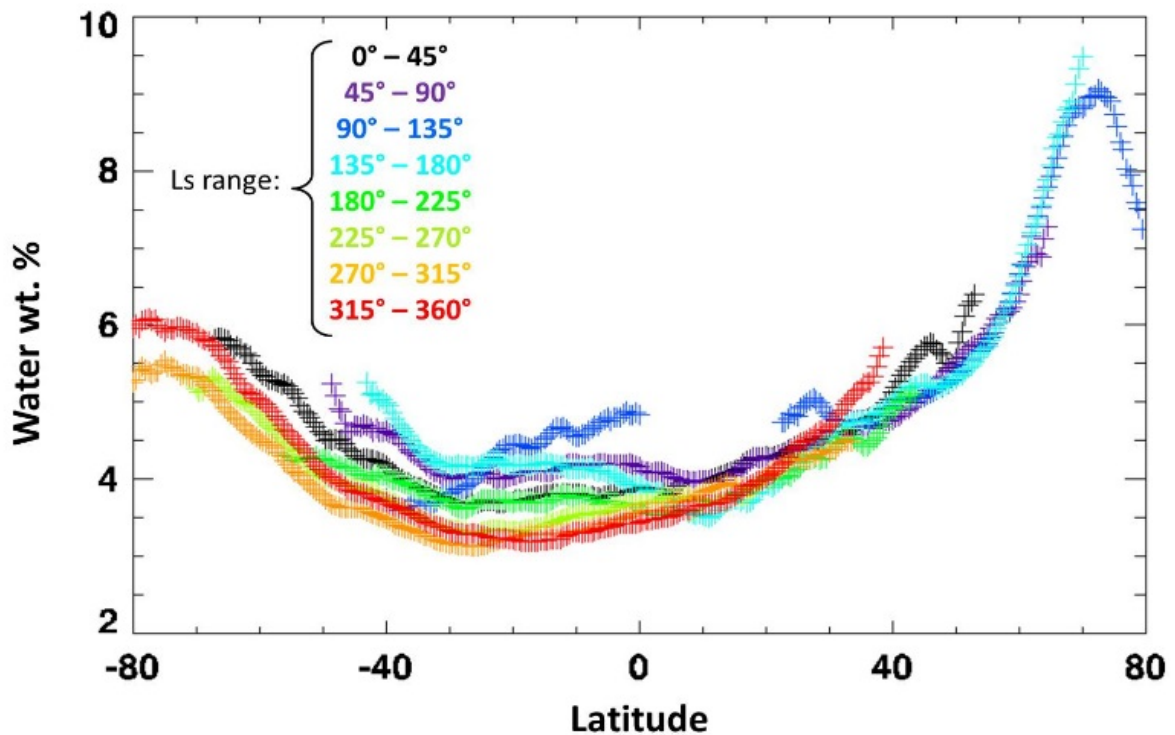

*Figure 8. Variation of Mars surface mean water wt. % as a function of latitude and for different seasonal range, indicated by different colors. To construct this diagram, we selected only data with water ice clouds index superior to 0.84 and τ    <2  Nevertheless, some water ice clouds between 20°S and 0° of latitude at Ls 90-135° are still present (dark blue crosses). The dependence of optical surface hydration to latitude is visible, and the northern high latitudes are more hydrated than the southern high latitudes. Apparent seasonal variations of hydration are addressed in section 3.2.*

## 3.2 Time variations

### 3.2.2 Diurnal variations

The orbit of the Mars Express spacecraft is elliptical, thus the present OMEGA dataset includes observations of a given location at multiple local times. Overall, there is no unambiguous correlation between the local time of the acquisition and the water wt. %. Some areas near the North Pole are even observed twice during the same sol but no diurnal variation of surface hydration was observed.

### 3.2.1 Seasonal variations: update view of OMEGA.

As shown in Figure 8, our mapping does not exhibit any clear seasonal variation of hydration in the northern hemisphere for latitudes > 40°N, in contrast to previous analyses using OMEGA data [*Milliken et al.*, 2007 ; *Jouglet et al.*, 2007   ]. After careful re-analyses of data, we consider that the reports of these variations by *Milliken et al.* [2007] was likely caused by unfiltered water ice clouds and that of *Jouglet et al.* [2007] possibly caused either by water ice clouds in the atmosphere or by water ice at the surface during the early spring observation (Ls 18° at a latitude of 50°N). Conversely, the seasonal curves of Figure 8 for latitudes lower than ~20°N are not well superimposed. To investigate these apparent seasonal variations, we focus on latitudes between 30°S and 0° where the seasonal coverage is complete and where the relative variations of hydration are large. As we suspect that these variations could be due to dust opacity variations, we plot the water content of theses latitudes as a function of their corresponding dust opacities $\tau_{eff}$ (Figure 9). It appears that $\tau_{eff}$ and water wt. % are anti-correlated, so that the southern seasonal variations of Figure 8 are best explained by variable quantity of aerosols in the atmosphere. Our analysis does not necessarily imply that the surface water content is constant with season, but potential seasonal variations are likely hidden by varying dust opacity and associated uncertainties (<~1.5 wt. %).

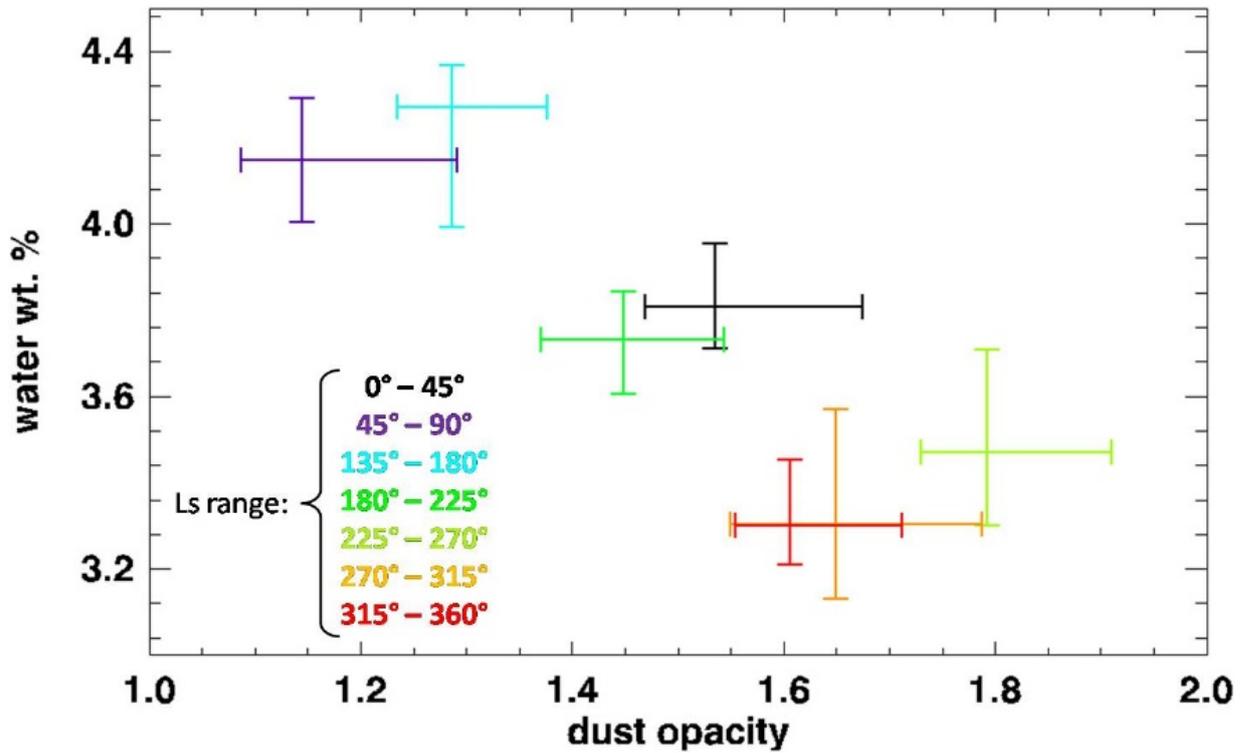

*Figure 9. Mean water wt. % as a function of mean dust opacity ($\tau_{eff}$) for data between 30°S and 0° and for different seasons (indicated by different colors). The error bars represent the latitudinal variability over the latitude range (according to a 1° sampling like in Figure 8). The Ls range 90° to 135° was excluded because of likely residual water ice clouds contamination at those latitudes.*

### 3.3 Comparison with in situ measurements

*Table 1. OMEGA observations of Gale crater northern plains, near Curiosity landing site (~900 × 900 m pixel).*

| OMEGA cube | Dust opacity | Water Ice clouds index | SNR | Phase angle | Ls | Temperature (K) | water wt. % | Δwater wt. % caused by SNR |
|---|---|---|---|---|---|---|---|---|
| **1577_3** | 2.15 | 0.78 | 200 | 44.4 | 190.0 | 252.9 | 5.02 | 0.24 |
| **2363_4** | 2.30 | 0.83 | 145 | 17.1 | 324.4 | 266.2 | 3.50 | 0.18 |
| **5273_4** | 1.80 | 0.83 | 128 | 50.2 | 29.5 | 243.1 | 4.68 | 0.26 |
| **6433_2** | 2.41 | 0.78 | 194 | 46.6 | 185.9 | 266.0 | 4.06 | 0.20 |
| **6676_4** | 2.83 | 0.84 | 220 | 13.9 | 227.9 | 277.2 | 2.87 | 0.13 |

OMEGA observations of the northern plains of Gale crater (table 1) yield a mean water content of 4.0 wt. % with an uncertainty standard deviation of 1.0 wt. %. Two data cubes that provide high water wt. % (ORB1577_3 and ORB6433_2) have an ice cloud index value close to the 0.8 threshold, which may indicate the presence of thin water ice clouds during the observations. On the other hand, data cubes ORB6676_4, ORB2363_4 and ORB5273_4 appear to be free of ice clouds and variation of water wt. % between cubes is consistent with their respective dust opacity (i.e. water wt. % decreases with increasing $\tau_{eff}$, as discussed in Section 2.3). Therefore, we consider that the data cube ORB5273_4, which corresponds to the lowest $\tau_{eff}$, provides the most reliable estimate of surface hydration of the northern plains within Gale, namely a water content of 4.7 ± 0.3 wt. %, with a possible 1 σ compositional bias of ~1.15 wt. %. This orbital estimate is larger than the Curiosity in situ measurements performed by the SAM instrument of the Rocknest sand shadow material, which yield bulk water content values between 1.5 and 3 wt. % [*Leshlin et al.*, 2013]. Therefore, the water content of the Rocknest soil volume material is possibly not representative of the optical surface water content at the kilometer-scale of OMEGA pixels, which might be more impacted by the hydrogen-rich thin dust layer at the surface seen by ChemCam first shots.

The discrepancy between the in situ and orbital measurements is even larger for the Phoenix landing site. *Poulet et al.* [2010], using the method of *Jouglet et al.* [2007], reported that OMEGA measured a surface hydration of 10-11% in comparison to the bulk water content of ~2% measured for an ice-free scooping with the TEGA (Thermal and Evolved Gas Analyzer) instrument [*Smith et al.*, 2009]. Using the method described in this paper and a larger data set, reassessment of the hydration signal of 11 OMEGA observations of Phoenix frost-free surroundings yields to a mean water content of 8.9 wt % with a 3σ of ~2 wt %. Note that no variation of water wt. % with Ls or local time is observed.

For the Phoenix site, such a large and systematic discrepancy possibly represents true variation in hydration with depth, and it could imply a different physical nature of the optical surface of the regolith. Our estimate of grain size, for instance, may not be valid for the Phoenix site. A value of A as low as 0.6, which would correspond to coarse (>250 µm) grains, would decrease the OMEGA-derived water content by a factor of 6. However, the Phoenix close-up imagery and orbital spectral analysis did not reveal peculiar difference in terms of grain size and composition, and the landing site regolith properties are consistent with the presence of typical Martian ferric dust [*Goetz et al.*, 2010; *Poulet et al.*, 2010]. As proposed by Poulet et al. [2010], the Phoenix and OMEGA results could be reconciled if the large concentration of $H_2O$ observed from orbit is confined to the few first millimetres or micrometers, whereas TEGA measured the water present in samples acquired from a depth of several centimeters. The dusty top µm of the surface near Phoenix landing site would then be two times more hydrated than the dusty surface at Gale crater. The fact that Phoenix TEGA and MSL SAM experiments measured similar bulk water wt. % values may indicate that if the top µm of the regolith as observed by OMEGA is much more hydrated at the high latitudes, the bulk is not. This leads us to conclude that the increase of hydration with latitude seen by OMEGA is driven by a surface-only process.

## 4. Origin and nature of water seen by OMEGA

### 4.1 Relation with GRS hydrogen

The neutron absorption measurements acquired by the Gama Ray Spectrometer (GRS) suite onboard Mars Odyssey [*Maurice et al.*, 2011; *Feldman et al.*, 2011] revealed strong $H^+$ abundances within the upper <1m at southern and northern high latitudes. This signal has been interpreted as water ice buried at shallow depth, which has been confirmed by the in situ detection of water ice by the Phoenix lander [e.g. *Mellon et al.*, 2011]. Permafrost is also likely present at latitudes down to ~40°, as water ice has been excavated from meter depth by recent impact at those latitudes [e.g. *Byrne et al.*, 2009]. It is important to note that the GRS and OMEGA instruments have very different sampling depths (<1 mm for OMEGA in comparison to several 10's cm for GRS). Therefore, the water content measured by the two instruments cannot be directly compared in a quantitative manner. On the other hand, water vapor diffusion between the buried ice and the atmosphere is expected and could (re)charge the regolith with water vapour that could be responsible for the increased water signature as seen by OMEGA.

However, several observations do not support the water diffusion as the origin of the OMEGA water signature. First, the low in situ estimate of water content at the Phoenix landing site is not consistent with the fact that the ground water ice/atmosphere transport mechanism could produce highly hydrated regolith [*Smith et al.*, 2009; *Poulet et al.*, 2010]. Second, water-equivalent hydrogen (WEH) maps derived from GRS neutron counts [*Maurice et al.*, 2011] show an increase with latitude with values that are very similar between the two hemispheres, which is not the case for the asymmetric trend of the surface hydration revealed by OMEGA (Figure 8). Finally, *Jakosky et al.* [2005] proposed that if some of the buried ice was not stable, its diffusion back to the atmosphere would produce an inter-annual variation in hydration of the top surface (i.e., a non-null water flux). We do not observe any clear indication of inter-annual variability of water content at the optical surface (when compared at same latitude and $\tau_{eff}$) in the present analysis of OMEGA data. Variations in hydration of the top surface are below the detection limit, if not zero.

## 4.2 Relation with water vapor

It has been suggested that part or all of the 3µm water absorption was the result of water adsorbed on particle surfaces, which is expected to exchange with the atmosphere over seasonal and possibly diurnal cycles. We have shown above that we do not detect significant variations of the amount of water as a function of local time and season. To further study possible exchange with the atmosphere, we first look for a relationship between the atmospheric relative humidity (RH, which is $P_{H_2O}/P_{saturation}$) and the observed surface water. For this purpose, we use a 3D Global Climate Model (GCM) [*Forget et al.*, 1999] to predict the RH of the atmosphere near the surface. RH varies with atmospheric water content and temperature, thus with location, season and local time. The GCM includes the effect of local slopes [*Spiga and Forget*, 2008], cloud microphysics [*Navarro et al.*, 2013] and radiative active clouds [*Madeleine et al.*, 2012]. The simulation is performed at a 2 pixel per degree resolution with a temporal resolution of one half of a Martian hour, using published TES albedo and thermal inertia maps [*Putzig and Mellon*, 2007] and local slopes and azimuth derived from MOLA.

For every OMEGA acquisition, we interpolate predicted RH as a function of longitude, latitude, and time (Ls and local time). The correlation between RH and water wt. % is presented in Figure 10. Some latitudes seem to exhibit a positive correlation with RH (-50 °N to -10°N), but these apparent variations may be a consequence of changing dust opacity (see bottom right plot of Figure 10). We have performed the same cross-correlations for various ranges in albedo and have achieved similar results. As a consequence, we do not observe any conclusive evidence of correlation between surface water wt. % and predicted RH for any latitude or type of surface (albedo). At low latitudes and over one Martian year, RH spans more than 4 orders of magnitude (from ~0.1% to 100%), whereas the OMEGA-based water wt. % of the surface remains constant except for small variations due to seasonally changing dust opacity.

The absence of diurnal and seasonal variations of surface hydration by OMEGA is consistent with in situ ChemCam experiments, which did not measure any significant (>1%) diurnal variation of the surface hydrogen signal [*Meslin et al.*, 2013]. Similarly, no correlation between water wt. % and local RH was observed when averaging over 1, 3 and 6 Martian months. If these observations are correct, it would imply that the bulk surface materials on Mars do not actively cycle water with the atmosphere on diurnal or seasonal cycles under current climatic conditions. However, hydrated minerals that are known to respond quickly to changes in RH may still be present in small quantities and below the detection limits of the methods used here. The low latitudes maintain a hydration of ~4 water wt. % that is apparently not exchangeable (variations < ~1 wt %), indicating that kinetics are too slow at the low temperatures making any hydrated minerals stable over these timescales, or that water is held tightly in materials and not responsive to changes in RH. The latter interpretation is consistent with water evolved at high temperatures in the Rocknest soil as measured by Curiosity, which may be in the fine-grained amorphous component [*Leshin et al.*, 2013].

Ice-free surfaces at high latitudes are only observed during their respective summer months. During these seasons, the RH does not vary much with latitude whereas the hydration increases with latitude. During the northern summer, the atmospheric RH at high latitudes is comparable to values achieved at low latitudes during other seasons. Despite this, the northern latitudes are much more hydrated during the summer and other seasons than ever experienced at lower latitudes. This indicates that appreciable amounts of water do not readily exchange with the atmosphere at the daily to seasonal scale under current climatic conditions. Therefore, other processes or latitudinal variations in regolith composition must be responsible for the increase and apparent sequestration of surface hydration at high latitudes. RH and surface temperature are at their highest values during summer and should not allow important quantities of adsorbed water to persist.

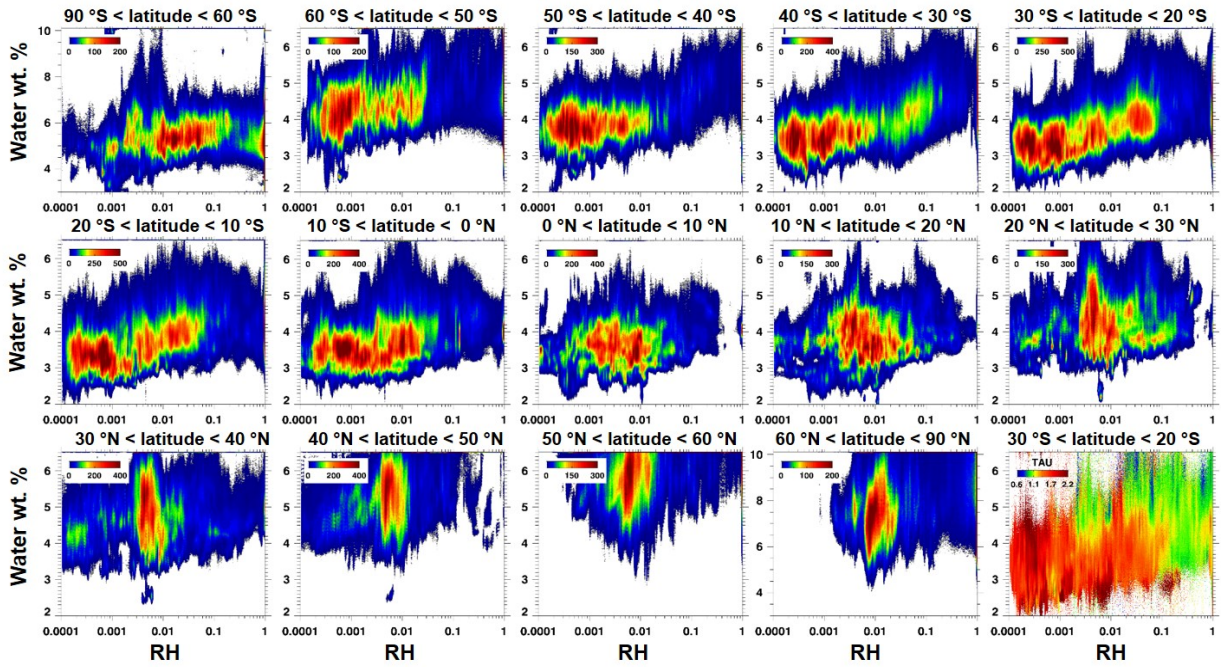

*Figure 10. Correlations of water wt. % values derived from OMEGA with RH predicted every hour. The latitudinal range of every correlation is indicated on top of the diagrams. The southern and northern high latitudes have a 30° of latitude range and the mid latitudes have a 10° of latitude range. The bottom right diagram shows the $\tau_{eff}$ value for the data composing the [-30°N, -20°N] diagram. The possible variations of Mars surface apparent hydration with RH (latitudes between -50°N and -10°N) can be interpreted as being caused by $\tau_{eff}$.*

To address potential links between the optical surface hydration and atmospheric water at longer timescales, we present in Figure 11 a map of the annual average of RH (denoted as $\overline{RH}$) and a map of the annual average of atmospheric $P_{H2O}$. The fact that higher altitudes in the southern hemisphere produce a reduced column of water is compensated by lower temperatures, yielding very similar $\overline{RH}$ for the two hemispheres. $\overline{RH}=1$ means that atmospheric water vapour is at its saturation level during all the year. $P_{H2O}$ and RH effectively trace the annual evolution of surface frost and ice. When RH is equal to 1, water frost is deposited at the surface, but only if $P_{H2O}$ is sufficient. For instance, a water frost annulus is observed during northern winter around the north pole $CO_2$ cap but despite similar RH at the southern high latitudes during southern winter, a water frost annulus is not systematically observed around the south pole $CO_2$ cap, where $P_{H2O}$ is much lower [*Kieffer et al.*, 2000; *Appéré et al.*, 2011; *Titus and Cushing*, 2014].

It is the integration of Figures 11a and 11b that allows a comprehensive visualization of the annual asymmetric deposition of water ice frost on Mars. In this context, the significant asymmetry in hydration levels between the two hemispheres appears correlated with the asymmetric repeated deposition of water frost at the surface. Although average RH is very high and hydration increases with latitude in both hemispheres, the estimated water content is much higher in the north. This may be because even though RH =1 near both poles, $P_{H2O}$ is higher in the north, leading to more water frost deposition. Such a correlation, if not fortuitous, could indicate that the seasonal presence of surface frost has a long-term effect on the hydration of surface materials. Indeed, $H_2O$ layers or films in a liquid-like state may exist between the regolith and the water frost [*Möhlmann*, 2008]. Although these layers would not be at equilibrium on a diurnal time scale (which is not readily observed by OMEGA), it is nevertheless possible that a long term effect water-regolith interaction could lead to stable increased hydration states at high latitudes. This could be fine-grained amorphous material (varnish-like) or it could be the long-term drive towards more hydrated minerals that are stable under these conditions. Though the specific mechanism is unclear, the data shown here reveal that the distribution of surfaces with the highest hydration state are correlated to those areas known to be subject to deposition of water frost under the present-day hydrologic cycle.

Alternatively, the different hydration levels between the southern and northern high latitudes could have two other explanations: (1) The RH predictions from the GCM do not correctly simulate the real (boundary) conditions at the surface of Mars; (2) the southern hemisphere regolith has different chemical or physical properties than the northern counterparts. Concerning (1), *Navarro et al.* [2013] report that the water cycle simulated by the GCM does not perfectly reproduce the water vapour column observed by TES [*Smith*, 2004] and OMEGA [*Maltagliati et al.*, 2011], especially at the tropics after the northern hemisphere summer (see section 5.3 of *Navarro et al.* [2013] for more details). A different vertical mixing of water vapour in the lower atmosphere than that parameterized in the model [e.g. *Maltagliati et al.*, 2013] might also modify the RH estimates near the surface. However, most of the observed Martian water cycle is reproduced well by the GCM simulation [e.g. *Montmessin et al.*, 2004]. Additional comparison with other simulations of the Martian water cycle should be performed to check if there is non-accuracy of the RH predictions discussed here.

Concerning (2), although the mineralogy of the southern high latitudes is one of the most difficult to constrain due to the long-term stability of $CO_2$ frosts and strong dust opacity, the spectral signatures do not exhibit particularly unusual or exotic properties. The spectral characteristics are actually quite similar to those of northern latitudes at regional or global scales [*Bandfield*, 2002; *Poulet et al.*, 2008]. The main difference between the two hemispheres is that the southern high latitudes are less dusty than in the north [*Ody et al.*, 2012]. Figure 5b shows that for the northern high latitudes (60-75 °N), the hydration increases with dust abundance at the surface. The positive correlation between hydration and dust abundance (at a given latitude) could imply that an important part of the high latitude hydration is caused by $H_2O$ adsorbed onto grain surfaces, because a surface with increased dust abundance will have a larger surface area per unit volume to interact with atmospheric water. Alternatively, the dust itself is likely to host a component of hydrated amorphous material [e.g., *Leshin et al.*, 2013]. Therefore, the hemispheric differences may be attributed to differences in mixing ratios between anhydrous basaltic materials and hydrous alteration components in fine-grained dust. Indeed, small variations in the presence of highly hydrated components, such as perchlorate or sulphate salts, may lead to increased absorptions in the strong fundamental absorptions in the 3 um region but not at shorter wavelength regions that host weaker overtone or combination vibration absorptions.

On the other hand, the difference in dust abundance between the two hemispheres is not quantitatively consistent with the observed hydration discrepancy between northern and southern high latitudes. Therefore, the distribution of dust at the optical surface is certainly correlated to some of the longitudinal variation of hydration at constant (high) latitude but does not fully explain the differences in hydration between the two hemispheres. A latitudinal coupling of surface hydration with high RH and frost deposition of atmospheric water is considered as the main driver of increased hydration with latitude (Figure 9). This interaction may in turn lead to variations in the hydration state of phases in local or regional regolith materials, either by converting minerals to more hydrated forms that are stable under these conditions (as may happen for sulfates or perchlorates), or by creation of new hydrous alteration components through regolith-water interaction.

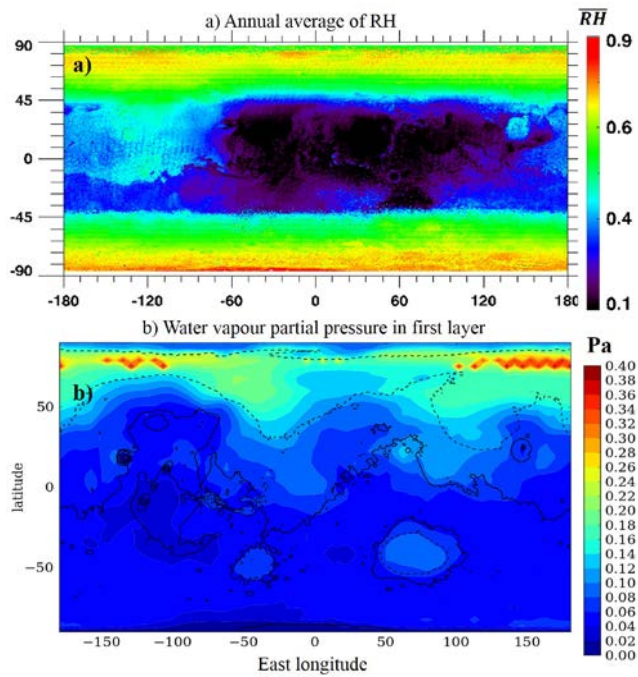

*Figure 11. Global maps predicted by the GCM model of a) $\overline{RH}$ (annual average of RH) and b) Annual average of water vapour partial pressure in first layer.*

## 4.3 Physical state of water in the Martian regolith

At the low latitudes of Mars, OMEGA does not observe any significant variation in hydration at the optical surface hydration as a function of particle size, RH, or $P_{H2O}$. Adsorbed water is expected to vary with these parameters [e.g. *Milliken and Mustard*, 2007b; *Pommerol and Schmitt*, 2008b; *Beck et al.*, 2010], thus it is unlikely that low-energy adsorbed water is responsible for the observed "<45 µm equivalent" 4±1 water wt. % constituting the background hydration state of the Martian optical surface.

In situ heating experiments of regolith materials by Curiosity and Phoenix measured lower water contents than what is inferred from orbital spectroscopy (1.5-3 wt. % and ~2 wt. % versus 4-5 wt % and 9-11 wt % respectively). However, the in situ samples were collected and sieved (<150 µm) prior to analysis. In addition, there was a time lag between sample scooping and analysis that exposed the materials to conditions different than what they had experienced when in place on the surface. This protocol could have preferentially removed some of the low-energy bound water in the samples. Because of the important uncertainties on the weight and potential destabilization of the sample prior to its deliver to SAM, Curiosity and OMEGA measurements of water wt. % may actually be in agreement within the uncertainties, such that the surface hydration as seen by OMEGA may be representative of the top centimeters of regolith at Gale. Such an interpretation is reinforced by the fact that the ChemCam experiment does not see any conclusive variation of the H signal with depth [*Meslin et al.*, 2013]. This gives additional constraints on the nature of the hydration detected by OMEGA. Most of the water vapour released from the four regolith samples of Rocknest occurred at temperatures ranging from ~350 K to 650 K [*Leshin et al.*, 2013]. The temperature of adsorbed water desorption depends on the composition of the mineral substrate and also on the microporous structure of the sample, but physisorbed water is expected to desorb below 350 K [*Dyar et al.*, 2010], whereas chemisorbed and structural water can resist much higher temperatures. Higher temperature water vapour release (T > ~500 K, though it also depends on the host material) is also consistent with loss of hydroxyl groups. It is likely that the 3 µm absorption observed in all Martian surface spectra is the result of both $H_2O$ and $OH^-$, perhaps in varying proportions, though the presence of atmospheric $CO_2$ precludes detailed analysis of fundamental $OH^-$ vibrations in the 2.65-2.85 µm wavelength region using orbital data. Considering the high temperatures at which most of the $H_2O$ release was measured during the SAM experiments, we conclude that most of the released water was not originally adsorbed water but consisted in structural and chemisorbed water molecules as well as re-combinative loss of structural hydroxyl.

The increased hydration that we observe at high latitudes does not appear to vary with RH under the frost-free surface conditions analyzed here, arguing against rapid diurnal or seasonal exchange of low-energy water under current climatic conditions. As seen previously, increased hydration at high northern latitudes is correlated to regions that experience deposition of water frost, possibly explaining latitudinal variations and the dichotomy between the two hemispheres in terms of bulk water content. The present hydration of the high latitudes might reveal an effect due to interaction with water ice/frost that is cumulative on a time scale greater than a single frost cycle. This does not exclude the possibility that the hydration can vary locally with dust abundance (and thus available surface area) at these latitudes, possibly because of enhanced interaction with atmospheric-deposited water frost at the surface. However, the similarity in water content between bright (dusty) and dark (less dusty) regions at equatorial and tropical latitudes indicates there is not a direct link between dust abundance and water content at a global scale; any such relationship must be mediated by properties and processes other than available surface area.

## Conclusion

We have built upon the previous work of *Milliken et al.* [2007] and *Jouglet et al.* [2007] to infer the water content (wt. %) of the Martian optical surface based on the ubiquitous 3 µm absorption as seen in OMEGA reflectance spectra. A more comprehensive data filtering and processing routine has been developed and employed to remove several biases present in these previous studies. In addition, the results presented here are based on a much longer time series (four Mars years) that allows us to explore seasonal variations in surface hydration in unprecedented temporal detail. The quantitative hydration estimates used in this study are based on a technique that rests upon detailed laboratory studies of numerous hydrous minerals and phases [*Milliken*, 2006]. However, this technique still remains strongly dependent on the particle size, such that the derived hydration values are dependent on the choice of model parameters. A global map of water wt. % at a 32 ppd spatial resolution is derived and apparent variations are discussed. It is established that estimated water content:

- varies with latitude, where regions poleward of ~45° exhibit a non-ambiguous North/South dichotomy in hydration levels; low latitudes (<30°) have a "background" hydration of 3-5 wt. % that increases to ~13 wt. % at the northern high latitudes and ~8 wt. % for the southern high latitudes;
- varies with the mineralogical composition of the surface, revealing the importance of structural water in minerals such as clays;
- is correlated with atmospheric components such as airborne dust and water ice clouds, which complicates the retrieval of hydration associated purely with surface properties.

OMEGA does not reveal any apparent diurnal or seasonal variation attributable to a physical change of surface hydration. Previously reported variations using OMEGA spectra were likely caused by incomplete removal of water ice clouds. Apparent variations are small (if not null) for all latitudes under frost-free conditions. Despite the relative humidity of the near-surface atmosphere spanning four orders of magnitude, the surface hydration remains stable throughout the Martian year. This implies that most of the $OH^-$ and $H_2O$ molecules that give rise to the 3 µm absorption are not exchangeable with the atmosphere on these timescales under current climatic conditions. This may be because much of the $H_2O$ is strongly bound to minerals/amorphous components of the Martian regolith or because kinetics of dehydration and rehydration are too slow under current martian temperature conditions. The former is consistent with the high temperature water vapour release observed in SAM experiments of regolith materials in Gale crater [*Leshin et al.*, 2013] and with the non-observation of a diurnal variation of the hydrogen signal with the LIBS ChemCam experiment [*Meslin et al.*, 2013]. Our best estimate of typical water contents in bulk regolith materials at equatorial and tropical latitudes is 3-5 wt. % $H_2O$, and these abundances are roughly constant for all longitudes at these latitudes, regardless of albedo and/or dust content. These values are in rough agreement with the SAM measurements when accounting for possible uncertainties. Regions known to contain large abundances of clay minerals or sulphate salts are exceptions to these 'background' values and exhibit higher water contents, whereas chloride-bearing deposits appear to exhibit lower water contents.

In contrast to equatorial regions, hydration at high latitudes is maximum where the dust abundance at the surface is greatest. Relative to the low latitudes, this suggests a primary control on the accumulation and sequestration of water in the regolith that is linked to but distinct from increased available surface area. The process is likely related to the seasonal deposition of water frost, possibly acting more efficiently where more dust is present (because of smaller grain sizes and hence higher available surface area for regolith-water interaction). This suggests the present-day surface hydration is stable under the current water cycle and Martian obliquity. The higher water content observed by OMEGA at high latitudes does not diminish during a summer or between winters, despite strong variations in RH, indicating the bulk of this reservoir of water is not exchangeable with the atmosphere under current climatic conditions.,

Future work will focus in integration of OMEGA data (acquired after Mars Express orbit #8485, for which no C channel data is available) as well as CRISM data (for which a specific thermal removal shall be implemented). Comparison of water wt. % values to other water vapour and relative humidity simulations will also be explored to confirm the results presented here. Additional in situ measurements by Curiosity and future landed missions will be also necessary to further constrain the origin of Martian surface hydration.

# Acknowledgements

Data from the OMEGA instrument and processing tools are available through the European Space Agency PSA FTP website. The authors would like to thank A. Pottier and F. Forget for providing water cycle simulations and together with P. Beck, A. Pommerol and P.-Y. Meslin for fruitful discussions concerning this work. Finally, we thank the two anonymous reviewers whose careful reviews improved this manuscript.